# On the effect of blockage objects in dense MIMO SWIPT networks

Ayse Ipek Akin, Ivan Stupia, and Luc Vandendorpe, *Fellow, IEEE*

## Abstract

Simultaneous information and power transfer (SWIPT) is characterised by the ambiguous role of multi-user interference. In short, the beneficial effect of multi-user interference on RF energy harvesting is obtained at the price of a reduced link capacity, thus originating nontrivial trade-offs between the achievable information rate and the harvestable energy. Arguably, in indoor environments, this trade-off might be affected by the propagation loss due to blockage objects like walls. Hence, a couple of fundamental questions arise. How much must the network elements be densified to counteract the blockage attenuation? Is blockage always detrimental on the achievable rate-energy trade-off? In this paper, we analyse the performance of an indoor multiple-input multiple-output (MIMO) SWIPT-enabled network in the attempt to shed a light of those questions. The effects of the obstacles are examined with the help of a stochastic approach in which energy transmitters (also referred to as power heads) are located by using a Poisson Point Process and walls are generated through a Manhattan Poisson Line Process. The stochastic behaviour of the signal attenuation and the multi-user interference is studied to obtain the Joint Complementary Cumulative Distribution Function (J-CCDF) of information rate and harvested power. Theoretical results are validated through Monte Carlo simulations. Eventually, the rate-energy trade-off is presented as a function of the frequency of walls to emphasise the cross-dependences between the deployment of the network elements and the topology of the venue.

A. I. Akin, I. Stupia and L. Vandendorpe are with the Institute of Information and Communication Technologies, Electronics and Applied Mathematics (ICTEAM), Université catholique de Louvain, Louvain la Neuve, Belgium.

## I. INTRODUCTION

*A. Background and motivations*

Simultaneous information and power transfer (SWIPT) is an emerging concept to increase the battery life of low power wireless devices. The main idea behind SWIPT is to use the same signal to transfer information and power simultaneously. The achievable trade-off between harvested energy and information rate in a single SWIPT link was originally presented by Varshney in [1]. Few years later, the authors of [2] showed the benefits of multi-input multi-output (MIMO) techniques on the performance of the SWIPT link with practical receiver architectures.

At the network level, particular interest in designing SWIPT systems is the ambivalent role of the multi-user interference (MUI). Traditionally, this is considered as an undesired factor for wireless information transfer (WIT) due to the negative effect on the coverage probability and the information rate [3], [4]. On the contrary, for wireless power transfer (WPT), the interference can be used as an additional source of energy to be harvested [5], [6]. It is doubtless that interference may have a major impact in determining the achievable rate-energy trade-off in SWIPT-enabled networks. MUI is even more crucial when the constraints on the minimum received power needed by the current RF harvesting technologies (tens of microwatts) are considered. These constraints, together with the limitations on RF emissions for guaranteeing the public safety, suggest that the deployment of SWIPT power heads (PHs) must be denser than that of traditional access points in wireless networks. Another straightforward consequence of the network elements' densification is that the performance of a SWIPT network is also highly dependent on the spatial locations of the network elements, thus making the Wyner model and the regular hexagonal or square grid models (see [7], [8]) inadequate to describe SWIPT systems whose PHs are likely to be randomly located. Recently, stochastic geometry has emerged as a random and spatial approach for modelling such kinds of dense networks, [9].

A first study that analyses the performance of SWIPT-enabled systems with a stochastic approach is reported in [10]. In this paper, the authors have proposed a tractable mathematical approach for the system-level analysis and optimization of SWIPT-enabled outdoor cellular networks. In [11], a mathematical framework is presented for MIMO SWIPT-enabled outdoor cellular networks. Though the methodology that is presented in both papers is extremely valuable in providing a tool for analysing the performance of stochastic SWIPT-enabled networks, their studies consider an outdoor cellular network that is too sparse to satisfy the constraints on the minimum received power that would enable RF energy harvesting. Moreover, whereas the distance is the determinant factor for outdoor wireless networks, propagation in indoor environments is also affected by the blockage due to the walls. Therefore, distance-dependent functions are not sufficient to model the propagation-losses for SWIPT indoor systems. In this perspective, a new stochastic geometry analysis for in-building systems, modelling the walls' distribution as a Manhattan Poisson Line Process (MPLP), [12], was presented in [13]. However, the assumption behind this work is that the blockage-based penetration loss is the dominant factor and free-space propagation loss can be neglected. This approximation conflicts with the typical operating conditions of SWIPT systems, in which the minimum received energy must be in the order of few microwatts to enable RF harvesting, so that a small difference on the distance can have a huge impact on the achievable rate-energy trade-off. More realistically, in [14], the authors jointly considered distance-dependent path loss, wall blockage and small-scale fading in information-only wireless networks by examining the average number of blockages for several wall generation methods. In this study, they considered different fixed transmitter configurations that provide the best scenario with respect to the interference. To the best of our knowledge, for SWIPT-enabled indoor networks, there has not been any study showing the effect of the obstacles combined with distance-dependent path-loss.

*B. Novelty and contributions*

All these concerns motivate us to propose an accurate stochastic geometry analysis for dense in-building MIMO power splitted SWIPT networks. MPLP is considered for the distribution of the walls and PHs are located by using Poisson Point Process (PPP). As a channel model, distance dependent path-loss, blockage-based path loss and fast fading are considered. In order to study the trade-off between information rate and harvested power, an analytical expression of the Joint Complementary Cumulative Distribution Function (J-CCDF) is derived by jointly considering the effects of obstacles, distance and multiple antennas. Eventually, our mathematical framework is validated through Monte Carlo simulations.

The main contributions of this paper can therefore be summarised as follows:

- In contrast to prior works relevant to SWIPT, [10], [11], we consider the effect of randomly located blockage objects like walls. This is of particular importance for any scenario accounting for the realistic levels of power enabling RF harvesting.
- Differently from [13], we consider the joint effect of the distance dependent propagation loss and the blockages. Moreover, the locations of both PHs and walls are modelled through a stochastic process. This allows a macro-level investigation of the interactions between the network and the venue in which it is deployed.
- An analytical expression of the cumulative distribution function of the multi-user interference as a function of the frequency of walls is provided. A similar result is also obtained for the distribution of the minimum propagation loss.
- The proposed numerical results provide important insights on the interplay between the network and the in-building environment from the perspective of the energy-rate trade-off. We also clarify the role of multiple antennas and SWIPT receiver architecture when an ultra-dense deployment of the network elements is considered.

The rest of this paper is organized as follows. Section II describes the system and path loss model. We focus on performance analyses in Section III. In Section IV, numerical results are presented and finally the paper is concluded in Section V.

*Notation:* $j = \sqrt{-1}$ denotes the imaginary unit. $\mathbb{E}(\cdot)$ is the expectation operator. $\mathbb{1}\{\cdot\}$ is the indicator function. $\text{Im}\{\cdot\}$ denotes the imaginary part. $\Gamma(\cdot,\cdot)$ is the upper-incomplete Gamma function [15, Eq. 8.350.2]. ${}_pF_q(a_1,...,a_p;b_1,...,b_q;\cdot)$ is the generalized hypergeometric function [15, Eq. 9.14.1]. $\mathcal{H}(\cdot)$ denotes the Heaviside function and $\bar{\mathcal{H}}(\cdot) = 1 - \mathcal{H}(\cdot)$.

## II. SYSTEM MODEL

In this section, we first introduce the major elements constituting a MIMO SWIPT network. Then, some blanket assumptions on the spatial distribution of both the network elements and the blockage objects are made clear. For illustrative purposes, an instance of indoor SWIPT network over a finite area is shown in Fig.1, where the lines represent walls and the markers symbolize PHs. It is worth recalling that this paper targets a stochastic characterisation of the SWIPT performance for a generic low power device (LPD) equipped with a power splitting receiver. Without loss of generality, in the remainder of the paper it is assumed that the LPD is conveniently located at the origin of the $x-$ and $y-$ axes.

### A. SWIPT Enabled Indoor Model

We consider a MIMO SWIPT network deployed in a two-dimensional finite indoor area. The SWIPT waveforms are generated through a set a randomly deployed PHs equipped with $n_t$ antennas. The LPD embedding a SWIPT receiver with $n_r$ antennas is located at the centre of the investigated region. It is assumed that the data delivered to the LPD comes from the PH providing the minimum average signal attenuation. In order to increase the performance of the information transfer process, the received signals captured from the different LPD antennas are processed using maximum ratio combining (MRC), while maximum ratio transmission (MRT) is implemented at the transmitter side. The SWIPT receiver operates according to the power splitting (PS) scheme: the received signal is split in two streams of different power levels by using a power splitting ratio $\rho$, where $0 \leq \rho \leq 1$. While one signal stream is sent to the rectenna circuit for energy harvesting, the other stream is used for information decoding.

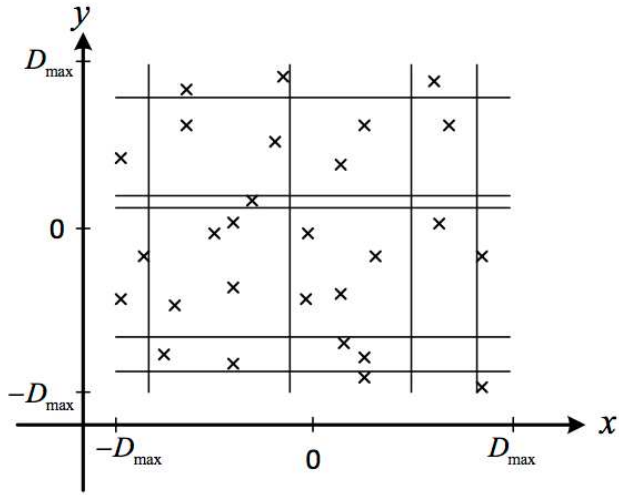

Fig. 1: One realization of the MPL and PP processes.

### B. Signal Propagation

Similarly to what has been done in [14], we assume that the signals are subject to distance dependent path loss, wall blockages and small scale fading. The path loss and the wall blockages are jointly modelled through the following log-distance dependent law:

$$l_N(r) = \frac{\kappa r^\beta}{K^N} \qquad (1)$$

where $r$ is the distance between the PH and the device, $\beta$ is the path-loss exponent, $K \in [0, 1)$ is the so called penetration loss, and $N$ is a random variable representing the number of walls between a generic PH and the device. The path loss constant is defined as $\kappa = (\frac{4\pi}{v})^2$, where $v = c_0/f_c$ is the transmission wavelength, $f_c$ and $c_0$ being the carrier frequency (Hz) and the speed of light (m/sec), respectively. The small scale fading is modelled through random variables following a Rayleigh distribution to account for the multi-path propagation effect.

## C. Spatial distribution of PHs

For analysing an ultra-dense MIMO SWIPT network in an indoor environment, we capitalise on a stochastic geometry approach. To achieve this, the possible sets of PHs deployed in a given area are modelled as instances of an homogeneous PPP $\Psi$ of density $\lambda_{PH}$. Thanks to this stochastic approach, we will obtain a statistical characterisation of the performance metrics for a typical LPD. As already mentioned, we assume that the information transfer towards the typical LPD is guaranteed by the PH ensuring the smallest average signal attenuation, also referred to as serving PH. From the information transfer perspective the other PHs are considered as interferers. The set of interfering PHs is denoted by $\Psi^{(\backslash 0)}$.

## D. Random wall placement

Differently from previous studies (e.g. [11]), this paper focuses on an indoor environment. It follows that our signal propagation model must comprise wall blockages. Again, we adopt a stochastic approach in which the position of wall is modelled as Manhattan Poisson Line Processes (MPLP). Generally speaking, MPLP is used to generate random lines in an Euclidean space $\mathbb{R}^n$. For the 2-D case, this is achieved by defining two homogeneous PPPs $\Psi_x$ and $\Psi_y$, of identical frequency $\lambda_w$, over the x-axis and the y-axis, respectively. Each set of points is obtained as a single realisation of $\Psi_x$ and $\Psi_y$ and it represents the midpoints of infinite length walls. Hence, the walls grow parallel to the x-axis and the y-axis at every point of these processes and each wall divides the plane into infinite rectangular boxes. Each of these boxes is considered as a room of our indoor environment. The room that contains the origin, called typical room, is identified by the couple $(0,0)$, while the other rooms are labelled according to their position with respect to the typical room, e.g. the signal associated with a PH located in the room $(i,j)$ shall cross $|i|$ walls on the x-axis and $|j|$ walls on the y-axis to reach the typical LPD. In order to justify our assumptions, we mention here the work presented in [14], wherein MPLP modelling is identified as the most promising wall generation method to approximate a realistic indoor scenario while guaranteeing a mathematically tractability of the wall blockages analysis.

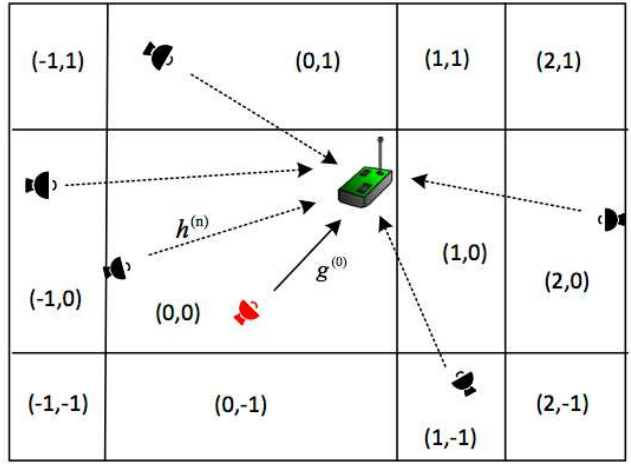

Fig. 2: SWIPT indoor network

## III. PERFORMANCE ANALYSIS

In this section we provide a methodology to study the effects of the blockage objects on the SWIPT performance. Then, we provide an expression for the J-CCDF of information rate and harvested power as a function of the network parameters. An illustration of a SWIPT network in an indoor scenario is shown in Fig. 2, where the serving PH (the red one in Fig. 2) is transmitting data and power towards the LPD with power gain $g^{(0)}$ (including MRT, MRC) and the signals, coming from all the PHs, experience Rayleigh fading.

## A. Wall Blockage Analysis

The main analytical contribution of this paper is the study of the effect of blockage objects on the SWIPT performance. To achieve this, we first decompose the original PPP, $\Psi$, with density $\lambda_{PH}$ into the sum of equivalent inhomogeneous PPPs, $\Psi_N$, with densities given by

$$\lambda_N(r,\theta) = \lambda_{PH} P_N(r,\theta) \qquad (2)$$

where

$$P_N(r,\theta) = \frac{(\lambda_w r|\cos(\theta)| + \lambda_w r|\sin(\theta)|)^N}{N!} \exp\{-(\lambda_w r|\cos(\theta)| + \lambda_w r|\sin(\theta)|)\} \qquad (3)$$

is the probability that a PH located at the point defined by the polar coordinates $(r,\theta)$ experiences the blockage effect of $N$ obstacles.

Since the choice of the serving PH is associated with the geometry of both the PHs' spatial distribution and the placement of the walls, the preliminary step enabling the analysis of the performance for the system depicted in Fig. 2 is the characterisation of the stochastic behaviour of the minimum path-loss, $L^{(0)}$, and the multi-user interference, $\mathcal{I}_{MU}$. This will be the objective of the following subsections.

*1) Minimum Path Loss:* As it has been already mentioned, we assume that the serving PH is the one associated with the minimum path loss, i.e.

$$L^{(0)} = \min_N \left\{ \min_{n \in \Psi_N} \{l_N(r^{(n)})\} \right\} \qquad (4)$$

where $n$ is the index of the PHs belonging to the inhomogeneous PPPs associated with $N$ obstacles, $\Psi_N$. Following the procedure proposed in [9], we capitalise on the well-known displacement theorem to end up with the following proposition.

*Proposition 1:* Let the minimum path loss being the one defined in (4) and define a function $\chi_\eta(\lambda_w)$ as

$$\chi_\eta(\lambda_w) = \lambda_w^\eta \left[ \frac{2^{\frac{\eta}{2}} \sqrt{\pi}\, \Gamma(\frac{\eta+1}{2})}{\Gamma(\frac{\eta+2}{2})} - \frac{\sqrt{2}\, {}_2F_1\left(\frac{1}{2}, \frac{\eta+1}{2}, \frac{\eta+3}{2}, \frac{1}{2}\right)}{\eta+1} \right] \qquad (5)$$

where $\Gamma(\cdot,\cdot)$ is the upper-incomplete Gamma function and $_2F_1(\cdot)$ is the Gaussian hypergeometric function [15]. Then the cumulative distribution function (CDF) of $L^{(0)}$ is given by

$$F_{L^{(0)}}(\alpha) = \Pr\{L^{(0)} \leq \alpha\} = 1 - \prod_{N=1}^{N_{max}} \exp\{-\Lambda_N([0,\alpha))\} \qquad (6)$$

where $N_{max}$ is the maximum number of obstacles that can be encountered in a circular region of ray $R_D$ and

$$\Lambda_N([0,\alpha)) = \begin{cases} \frac{4\lambda_{PH}}{N!} \sum_{\eta=N}^{\infty} \frac{(-1)^{\eta-N}\chi_\eta(\lambda_w)}{(\eta-N)!(\eta+2)} \left(\frac{\alpha K^N}{\kappa}\right)^{\frac{\eta+2}{\beta}} & \text{if } \alpha < \frac{R_D^\beta \kappa}{K^N} \\ \frac{4\lambda_{PH}}{N!} \sum_{\eta=N}^{\infty} \frac{(-1)^{\eta-N}\chi_\eta(\lambda_w)}{(\eta-N)!(\eta+2)} R_D^{\eta+2} & \text{if } \alpha \geq \frac{R_D^\beta \kappa}{K^N} \end{cases} \qquad (7)$$

is the intensity of the process $L_N = \{l_N(r^{(n)}), n \in \Psi_N\}$.

*Proof:* See Appendix A. ∎

From (7), it is apparent that the signal attenuation is a process whose intensity can be expressed as the weighted sum of power functions of $\alpha$. Interestingly, the effects of the blockage due to the walls is summarised through the weights $\chi_\eta(\lambda_w)$ that can be computed offline and tabulated for given values of $\eta$ and $\lambda_w$ as shown in Table I. Since $\chi_\eta(\lambda_w)$ is a decreasing function of $\eta$, from (6) and (7), we can infer that the greater the number of obstacles $N$, the lower the impact of the blockage objects on the CDF of $L^{(0)}$. On the contrary, $\chi_\eta(\lambda_w)$ is an increasing function of $\lambda_w$ and, not surprisingly, the impact of the blockage objects on $F_{L^{(0)}}(\alpha)$ increases with $\lambda_w$.

*2) Multi-User Interference:* The normalized (with respect to 1W of transmit power) multi-user interference can be expressed as

$$\mathcal{I}_{MU} = \sum_{N=0}^{N_{max}} \sum_{n \in \Psi_N} \frac{h^{(n)}}{l_N(r^{(n)})} \mathbb{1}\{l_N(r^{(n)}) > L^{(0)}\} \qquad (8)$$

where $h^{(n)}$ is an exponentially distributed random variable with unit variance representing the gain of the $n$th interfering link, and $r^{(n)}$ denotes the distance from a generic PH to the LPD.

| $\lambda_w \backslash \eta$ | 0 | 1 | 2 | 3 | 4 | 5 |
|---|---|---|---|---|---|---|
| 0.01 | 1.5708 | 0.02 | $2.5 \times 10^{-4}$ | $3.3 \times 10^{-6}$ | $4.35 \times 10^{-8}$ | $5.73 \times 10^{-10}$ |
| 0.02 | 1.5708 | 0.04 | $1 \times 10^{-3}$ | $2.6 \times 10^{-5}$ | $6.96 \times 10^{-7}$ | $1.83 \times 10^{-8}$ |
| 0.03 | 1.5708 | 0.06 | $2.3 \times 10^{-3}$ | $9 \times 10^{-5}$ | $3.5 \times 10^{-6}$ | $1.39 \times 10^{-7}$ |
| 0.04 | 1.5708 | 0.08 | $4.1 \times 10^{-3}$ | $2.1 \times 10^{-4}$ | $1.1 \times 10^{-5}$ | $5.87 \times 10^{-7}$ |
| 0.05 | 1.5708 | 0.1 | $6.4 \times 10^{-3}$ | $4.1 \times 10^{-4}$ | $2.7 \times 10^{-5}$ | $1.79 \times 10^{-6}$ |

TABLE I: Values of $\chi_\eta(\lambda_w)$

*Proposition 2:* Assume that the minimum path loss is given and equal to $L^{(0)}$, consider the functions $\chi_\eta(\lambda_w)$, $\eta \in \mathbb{R}_+$ as in (5), and define the function

$$\Delta_{\eta,N}\left(\omega; L^{(0)}\right) = \left(\frac{L^{(0)} K^N}{\kappa}\right)^{\frac{\eta+2}{\beta}} \left(1 -{}_2F_1\left(1, -\frac{\eta+2}{\beta}, 1 - \frac{(\eta+2)}{\beta}, \frac{j\omega}{L^{(0)}}\right)\right) \\ - R_D^{\eta+2}\left(1 -{}_2F_1\left(1, -\frac{\eta+2}{\beta}, 1 - \frac{(\eta+2)}{\beta}, \frac{j\omega K^N}{R_D^\beta \kappa}\right)\right). \quad (9)$$

Then, capitalizing on the Gil-Pelaez inversion theorem, [16], the CDF of $\mathcal{I}_{MU}$ is given by

$$F_{\mathcal{I}_{MU}}(z; L^{(0)}) = \Pr\{\mathcal{I}_{MU} \leq z | L^{(0)}\} = 1/2 - \int_0^\infty \frac{1}{\pi\omega} \operatorname{Im}\left\{\mathrm{e}^{-j\omega z} \prod_{N=1}^{N_{max}} \Phi_N\left(\omega; L^{(0)}\right)\right\} d\omega \quad (10)$$

where

$$\Phi_N\left(\omega; L^{(0)}\right) = \begin{cases} \exp\left\{\frac{4\lambda_{PH}}{N!} \sum_{\eta=N}^\infty \frac{(-1)^{\eta-N}\chi_\eta(\lambda_w)}{(\eta-N)!(\eta+2)} \Delta_{\eta,N}\left(\omega; L^{(0)}\right)\right\} & \text{if } L^{(0)} < \frac{R_D^\beta \kappa}{K^N} \\ 1 & \text{if } L^{(0)} \geq \frac{R_D^\beta \kappa}{K^N} \end{cases} \quad (11)$$

is the characteristic function of the multi-user interference produced by the PHs whose signals are subject to the attenuation of $N$ obstacles to reach the LPD.

*Proof:* See Appendix B. ■

## B. SWIPT Performance Analysis

The goal of this section is to provide an analytical expression for studying the stochastic behaviour of the SWIPT performance as a function of the PHs' density $\lambda_{PH}$ and the frequency of blockage objects $\lambda_w$. Because of the multiobjective nature of the system, two different performance metrics are considered, namely the average throughput, expressed in bits/sec and denoted with $R$, and the average harvested power, measured in Watt and referred to as $Q$. To achieve this goal, the instantaneous metrics are first defined as

$$R = B_c \log_2 \left(1 + \frac{Pg^{(0)}/L^{(0)}}{P\mathcal{I}_{MU} + \sigma_n^2 + \sigma_c^2/(1-\rho)}\right)$$
$$Q = \rho \xi P \left(\frac{g^{(0)}}{L^{(0)}} + \mathcal{I}_{MU}\right) \tag{12}$$

where $B_c$ is the signal bandwidth, $P$ is the average transmit power, $\rho$ is the power splitting ratio and $\xi$ is the energy harvesting efficiency factor. The variance of the thermal noise is denoted by $\sigma_n^2$, while $\sigma_c^2$ indicates the variance of the noise due to the conversion of the received signal from radio frequency to baseband. In addition to the minimum path loss $L^{(0)}$ and the multi-user interference $\mathcal{I}_{MU}$, whose statistical properties have been studied in the previous subsections, the other source of randomness is the power gain $g^{(0)}$, which encompasses the small scale fading experienced by the serving PH's signal and the effect of both the MRT and the MRC processing at the transmitter and the receiver side, respectively. The PDF of $g^{(0)}$ is given in [17] as,

$$f_{g^{(0)}}(\zeta) = \mathcal{K}_{m,n} \sum_{s=1}^{m} \sum_{t=n-m}^{(n+m-2s)s} a_{s,t} \zeta^t \exp(-s\zeta) \tag{13}$$

where $m = \min(n_t, n_r)$ and $n = \max(n_t, n_r)$. Here $\mathcal{K}_{m,n} = \prod_{k=1}^{m}((m-k)!(n-k)!)^{-1}$ is a normalising factor and $a_{s,t}$ are some coefficients that can be easily obtained by using [17, Algorithm 1].

In SWIPT enabled networks the performance of the system can be described in terms of achievable trade-offs between the information rate and the harvested power. This trade-off will be analysed taking advantage of the J-CCDF of $R$ and $Q$, as originally proposed in [10]. The J-CCDF is defined as

$$F_c(R^*, Q^*) = \Pr\{R \geq R^*, Q \geq Q^*\} \tag{14}$$

where $Q^* \geq 0$ is the sensitivity of the energy harvester and $R^* \geq 0$ is the minimum achievable rate. A convenient reformulation of the J-CCDF has been proposed in [11], and it amounts to computing the probability that the multi-user interference belongs to an interval for which a minimum SINR $\gamma$ can be achieved conditioned to a minimum amount of received power $q_*$, i.e.

$$F_c(R^*, Q^*) = \mathbb{E}_{L^{(0)}} \left\{ \int_{(\mathcal{T}_*/P)L^{(0)}}^{+\infty} F_{\mathcal{I}_{MU}}\left(x\gamma/L^{(0)} - \sigma_*^2/P \,\big|\, L^{(0)}\right) f_{g^{(0)}}(x) dx \right\} \\ - \mathbb{E}_{L^{(0)}} \left\{ \int_{(\mathcal{T}_*/P)L^{(0)}}^{+\infty} F_{\mathcal{I}_{MU}}\left(-x/L^{(0)} + q_*/P \,\big|\, L^{(0)}\right) f_{g^{(0)}}(x) dx \right\} \quad (15)$$

where $\sigma_*^2 = \sigma_n^2 + \sigma_c^2(1-\rho)^{-1}$, $q_* = Q^*/(\rho\xi)$, $\mathcal{T}_* = (q_* + \sigma_*^2)/(\gamma+1)$ and $\gamma = 1/\left(2^{R^*/B_c} - 1\right)$.

*Proposition 3:* Let define,

$$\Phi\left(\omega; L^{(0)}\right) = \prod_{N=0}^{\infty} \Phi_N\left(\omega; L^{(0)}\right), \\ \Lambda([0, \alpha)) = \sum_{N=0}^{\infty} \Lambda_N([0, \alpha)), \quad (16)$$

and $\widehat{\Lambda}([0, \alpha))$ as the derivative of $\Lambda([0, \alpha))$ with respect to $\alpha$, whose expression is provided in (27) in the Appendix B.

Given the statistical characterisation of $L^{(0)}$, $\mathcal{I}_{MU}$, and $g^{(0)}$ expressed as in equations (6), (29), and (13), respectively, the J-CCDF $F_c(R^*, Q^*)$ can be computed as [11],

$$F_c(R^*, Q^*) = \mathcal{K}_{m,n} \sum_{s=1}^{m} \sum_{t=n-m}^{(n+m-2s)s} a_{s,t} \left(J_{s,t}^{(1)} - J_{s,t}^{(2)}\right) \quad (17)$$

where the functions $J_{s,t}^{(1)}$ and $J_{s,t}^{(2)}$ are defined as,

$$J_{s,t}^{(1)} = \int_0^\infty \int_0^\infty \frac{1}{\pi w} \operatorname{Im}\left\{ \exp\left(-jw\frac{q_*}{P}\right) \left(s - \frac{jw}{y}\right)^{-(1+t)} \right. \\ \left. \Gamma\left(1+t, \frac{\mathcal{T}_*}{P}(sy - jw)\right) \Phi(w; y) \right\} \left(\widehat{\Lambda}([0,\alpha)) \exp\left\{-\Lambda([0,\alpha))\right\}\right) dw dy, \quad (18)$$

and

$$J_{s,t}^{(2)} = \int_0^\infty \int_0^\infty \frac{1}{\pi w} \text{Im} \left\{ \exp\left(jw\frac{\sigma_*^2}{P}\right) \left(s + \frac{jw\gamma}{y}\right)^{-(1+t)} \right.$$
$$\left. \Gamma\left(1+t, \frac{\mathcal{T}_*}{P}(sy + jw\gamma)\right) \Phi(w;y) \right\} \left(\widehat{\Lambda}([0,\alpha)) \exp\left\{-\Lambda([0,\alpha))\right\}\right) dw dy. \quad (19)$$

*Proof:* The proof trivially follows from [11, Proposition 1]. ∎

Using Proposition 3 the J-CCDF can be numerically computed for illustrating the average trade-off between the information rate and the harvested power without the need of time consuming Monte Carlo simulations.

## IV. NUMERICAL RESULTS

In this section, we show the performance of an in-building SWIPT network by means of numerical results. The analytical findings presented in Section III will be validated through Monte Carlo simulations.

### A. Setup

We considered a circular area of radius $R_D = 60$m in which a set of PHs is transmitting with an average power of 1W, i.e. $P = 30$dBm. The SWIPT signals have a bandwidth of $B_c = 200$kHz centred around $f_c = 2.1$ GHz. The thermal noise has a variance $\sigma_n^2 = -174 + 10\log_{10}(B_c) + \mathcal{F}_n$, where $\mathcal{F}_n = 10$dB is the noise figure. The variance of the noise due to the RF to DC conversion is set to $\sigma_c^2 = -70$dBm. The PHs are assumed to be deployed with a density $\lambda_{PH} = 1/(\pi d_{PH}^2)$, where $d_{PH}$ is half of the average distance between PHs. The efficiency of the RF energy harvesting process is $\xi = 0.8$ and the path-loss exponent is $\beta = 2.5$. Unless otherwise stated, the power splitting factor is $\rho = 0.5$, the number of receive antennas is $n_r = 2$ and the number of transmit antennas is $n_t = 4$. For the chosen frequency, according to [18, Table 3], the penetration loss $K$ can be reasonably set to $-10$dB per crossed wall.

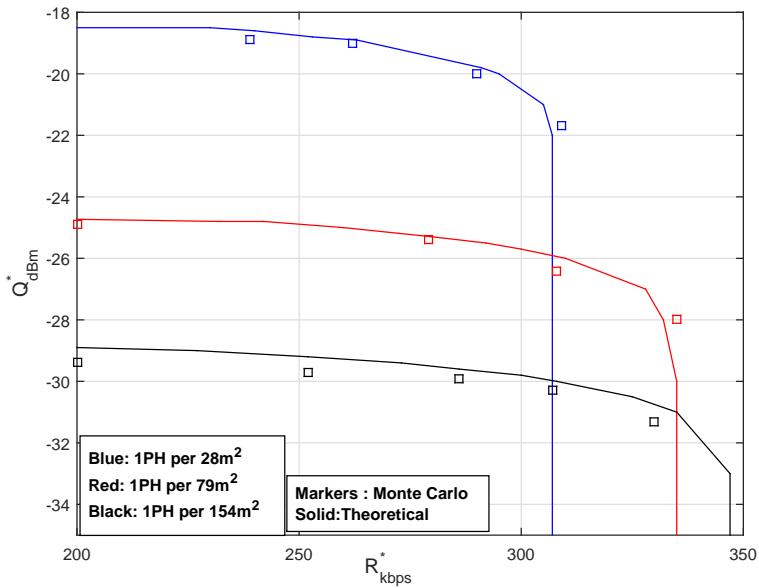

Fig. 3: Monte Carlo simulations and theoretical results of rate-energy trade-off for different $\lambda_{PH}$ when $\lambda_w = 0.05$, $n_t = 4$, $n_r = 2$.

## B. Results

*1) Validation of the analytical findings:* Fig. 3 shows the trade-off between the information rate and the harvested power parametrised with respect to $\lambda_{PH}$ when $F_c(R^*, Q^*) = 0.75$. The results are presented for different values of $d_{PH} = 3, 5, 7$ meters (i.e. $\lambda_{PH} = \frac{1}{28}, \frac{1}{79}, \frac{1}{154}$), while the frequency of the walls is kept fixed to $\lambda_w = 0.05$. The analytical results (solid lines) are compared to the ones obtained through Monte Carlo simulations (squares). Despite of the approximation due to the fact that the series in the equations (7) and (11) has to be cut at some point, the almost perfect match between theoretical results and simulations is apparent. Clearly, the harvested power always benefits from the increment of the PHs' density $\lambda_{PH}$, while the maximum achievable information rate decreases because of the larger level of multi-user interference. We can therefore identify a first compromise to be achieved when the network elements are densified. For example, if a minimum harvested energy of $-23$dBm is required, the optimum PHs' density is the one corresponding to $d_{PH} = 3$ meters (blue line), while if a harvested energy of $-28$dBm is sufficient, it would be better to have a less dense network, $d_{PH} = 5$ meters, to benefit of a higher information rate (red line).

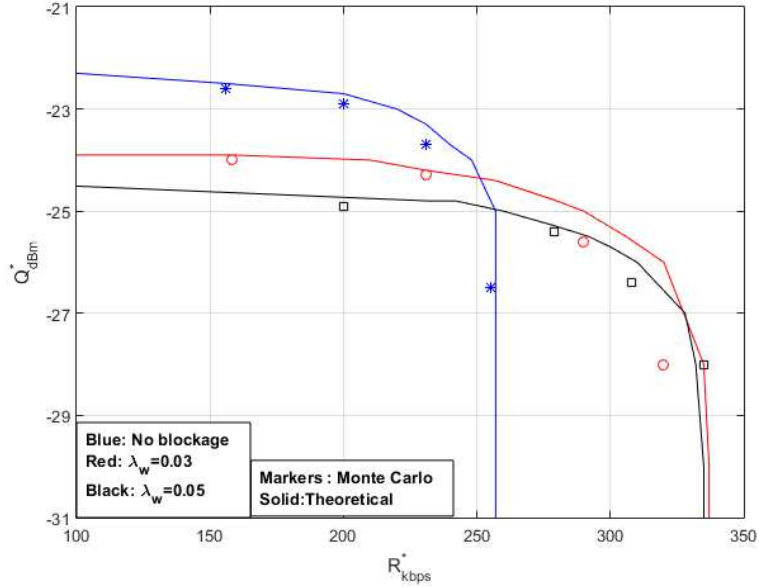

Fig. 4: Monte Carlo simulations and theoretical results of rate-energy trade-off for different $\lambda_w$ when $d_{PH} = 5m$, $n_t = 4$, $n_r = 2$.

In order to investigate the effect of the walls on the SWIPT performance, Monte Carlo simulations (markers) and theoretical results (solid lines) are provided in Fig. 4 when $F_c(R^*, Q^*) = 0.75$ and $d_{PH} = 5$ meters . The blue curve is used as a benchmark and represents the case in which all the blockage objects have been removed. The red and black curves are associated with two different values of $\lambda_w$, and specifically with $\lambda_w = 0.03$ and $\lambda_w = 0.05$, respectively. It is apparent that an increment of the frequency of blockage objects is beneficial for the maximum achievable information rate because of the reduced multi-user interference. On the other side, we notice a reduction of the harvestable power level when $\lambda_w$ increases. We can conclude, that the topology of the venue impacts the achievable rate-energy trade-off in view of the ambivalent role of the multi-user interference in SWIPT networks.

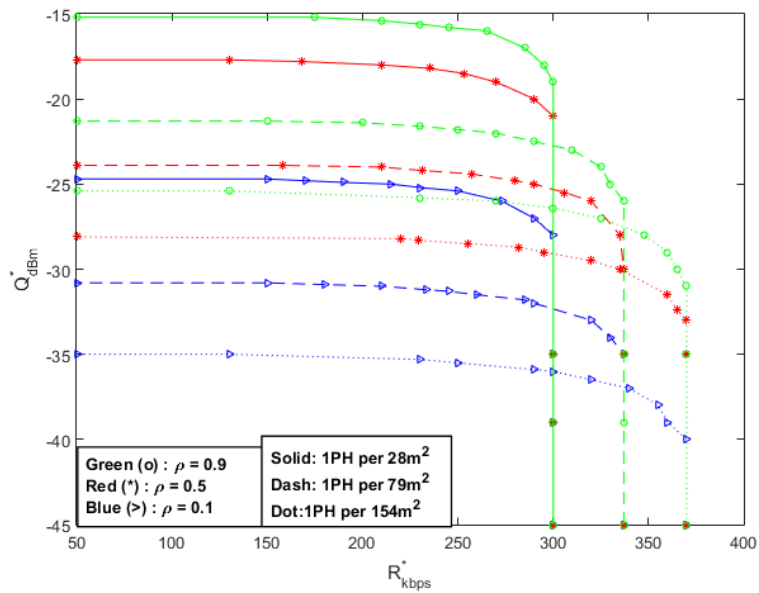

Fig. 5: Theoretical results of rate-energy trade-off for different $\rho$ values when $F_c(R^*, Q^*) = 0.75$, $\lambda_w = 0.03$, $n_t = 4$ and $n_r = 2$.

*2) SWIPT Receiver Analysis:* We now analyse the role of the SWIPT receiver architecture on the achievable performance. The rate-energy trade-offs for different values of the power splitting ratio when $F_c(R^*, Q^*) = 0.75$ are presented in Fig. 5. The curves are provided for 3 different values of $\lambda_{PH}$ when $\lambda_w = 0.03$. Obviously, the receiver harvests more power with higher power splitting ratio for all cases. More interestingly, we obtain the same value of maximum achievable information rate for all values of $\rho$. This is due to the fact that, with the level of densification required to receive a total power belonging to the microwatt region, the system is essentially interference limited and the SINR does not depend on $\rho$ for realistic level of noise. For instance, considering the case $d_{PH} = 3m$ (solid lines), with an information rate of 300 Kbps, $-27$dBm of harvested power can be achieved when $\rho = 0.1$, while we can obtain $-20$dBm and $-17$dBm of harvested power when $\rho$ is equal to $0.5$ and $0.9$, respectively. We can conclude that the power splitting ratio must be as large as possible and its fine tuning is of limited interest in the design of ultra-dense SWIPT networks. It is worth remarking that this conclusion is radically different from previous works (see [11]) in which the authors consider values of harvested power below $-60$dBm. In that case, multi-user interference has a limited role while the optimisation of the power splitting ratio is paramount. On the contrary, in this paper we show that the technological limitations of the harvesting process make the topology of the venue and its effect on the interference the most important parameters in defining the achievable rate-energy trade-off.

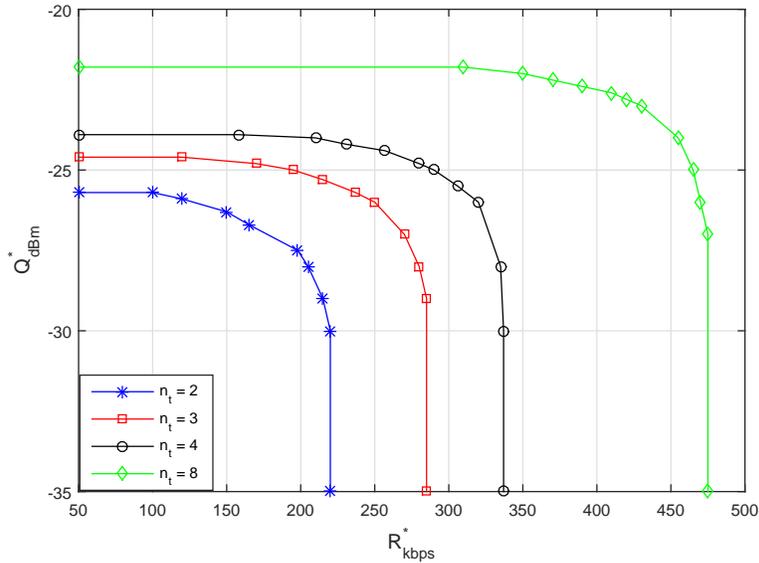

Fig. 6: Theoretical results for rate-energy trade-off as a function of $n_t$ when $n_r = 2$, $\lambda_w = 0.03$ and $d_{PH} = 5m$.

*3) MIMO Effect Analysis:* The previous subsections showed how the ambivalent role of the interference impacts the rate-energy trade-off. On the other side, both harvested energy and information rate can be improved by increasing the power gain from the serving PH to the LPD. This can be achieved by increasing the antenna array gain. In order to illustrate the benefits of MIMO on the SWIPT performance in such a dense interference-limited scenario, Fig. 6 provides the rate-energy trade-off for different numbers of transmit antennas when $F_c(R^*, Q^*)$ is 0.75 and $n_r = 2$. The proposed results show how increasing the number of antennas is of great help in dealing with the MUI while improving the level of harvestable power. Here, for illustrative purposes, we present numerical results only for $\lambda_w = 0.03$, but the same trend has been observed for all the tested frequencies of walls.

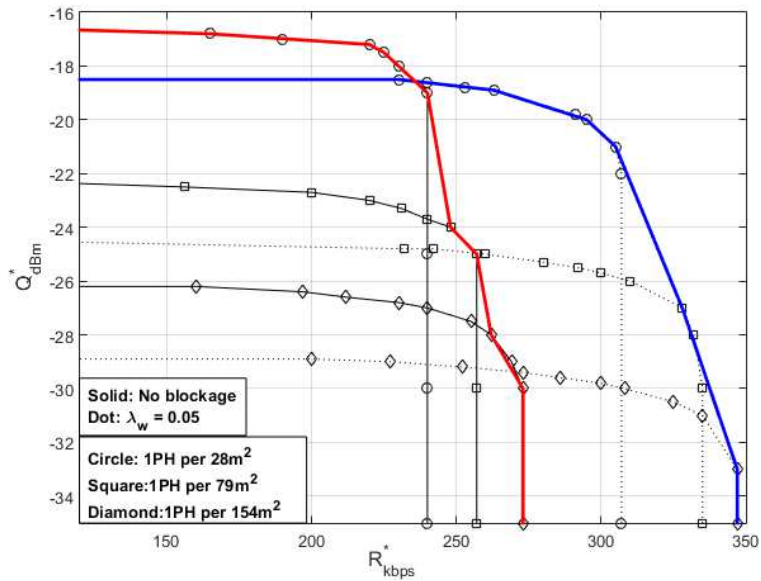

Fig. 7: Theoretical results for the envelope of rate-energy trade-off for different $\lambda_{PH}$ and $\lambda_w$ when $F_c(R^*, Q^*) = 0.75$, $n_t = 4$, $n_r = 2$.

*4) Joint effect of $\lambda_{PH}$ and $\lambda_w$ on the achievable rate-energy trade-off:* Eventually, we focus on how the the level of network densification must be wisely chosen in accordance with the topological characteristic of the indoor environment. To this purpose, Fig. 7 shows the envelopes of the rate-energy trade-offs relevant to different values of $\lambda_{PH}$ and for a J-CCDF equal to 0.75. In particular, the red curve represents a situation in which all blockage objects have been removed, while the blue curve is associated with the presence of walls with frequency equal to $\lambda_w = 0.05$. It is apparent that the additional attenuation of walls is beneficial for the achievable trade-off and rate. For example, an information rate of $300$ Kbps can be achieved with $-20$dBm of harvested power in the presence of blockages, while this value is reduced to $240$ kbps when those obstacles are removed. In Fig. 8 we plotted the energy-rate trade-off for different values of $\lambda_w$ with $d_{PH} = 3m$ and $d_{PH} = 7m$. In the denser case, the information rate always benefits from the reduced MUI due to the presence of walls. On the contrary, when the network elements are less dense, the blockage experienced by the serving PH is not compensated by the MUI reduction. Hence, neither the harvested power nor the information rate benefits from the interference. Hence, we can argue that the best achievable trade-off is a multidimensional factor in which the cross-dependences between the network infrastructure and the topology of the venue have a major impact.

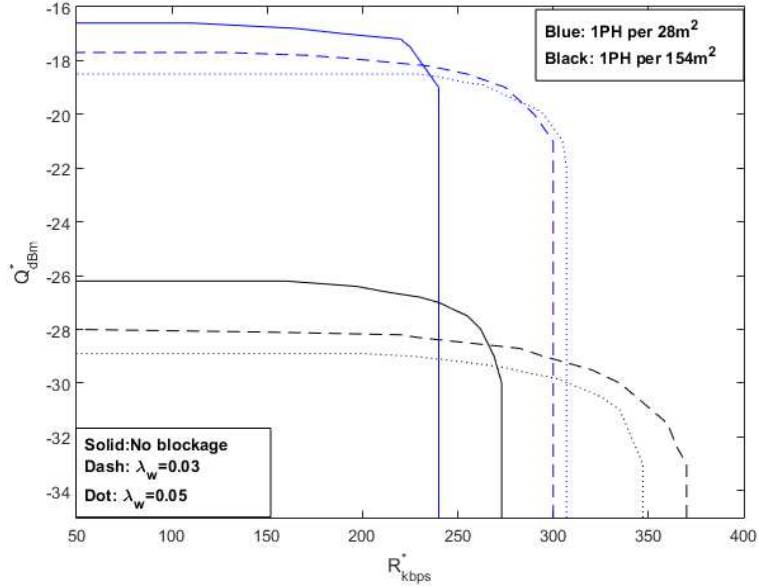

Fig. 8: Theoretical results of rate-energy trade-off for different $\lambda_{PH}$ and $\lambda_w$ when $F_c(R^*, Q^*) = 0.75$, $n_t = 4$ and $n_r = 2$.

## V. Concluding Remarks

In this paper, we considered a dense in-building MIMO SWIPT network by proposing an accurate stochastic geometry analysis. In this scenario, we addressed the issue of the MUI in relation with both the network infrastructure and the topology of the venue. We discovered that there is a nontrivial rate-energy trade-off between the PHs' density and the frequency of the obstacles. Blockage objects can be beneficial in some cases depending on which trade-off is considered. Their effect in reducing the received power is compensated by the decrement of the MUI. Moreover, we investigated the power splitting ratio and showed that it has no effect on the maximum achievable information rate while more power can be harvested with its higher values. Finally, it is demonstrated that multi antenna processing is a valuable strategy to counteract the effect of the densification. Our theoretical findings are validated through Monte Carlo simulations.

## Appendix A

### Proof of Proposition 1

First, we note that the process collecting all the propagation losses $L = \{l(r^{(n)}), n \in \Psi\}$ can be characterised as a transformation of the points of $\Psi$. Hence, invoking the displacement theorem, [9, Theorem 1.3.9], and recalling that $\Psi$ can be expressed as the superposition of the PPPs $\Psi_N$, the processes $L_N = \{l_N(r^{(n)}), n \in \Psi_N\}$, $N \in \{0, N_{max}\}$, are PPPs whose intensity is given by

$$\Lambda_N([0,\alpha)) = \Pr\left\{\frac{\kappa(r^{(n)})^\beta}{K^N} \in [0,\alpha), n \in \Psi_N\right\} \tag{20}$$
$$= \lambda_{PH} \int_0^{2\pi} \int_0^\infty \mathcal{H}\left(\alpha - \frac{\kappa r^\beta}{K^N}\right) P_N(r,\theta) \, r dr d\theta.$$

Now, by substituting (3) in (20) and integrating it over $r$, we get

$$\Lambda_N([0,\alpha)) = \frac{\lambda_{PH}}{N!} \int_0^{2\pi} (\lambda_w |\cos(\theta)| + \lambda_w |\sin(\theta)|)^{-2}$$

$$\times \left[ \left( \Gamma(2+N) - \Gamma\left(2+N, \left(\frac{\alpha K^N}{\kappa}\right)^{1/\beta} (\lambda_w |\cos(\theta)| + \lambda_w |\sin(\theta)|)\right) \right) \bar{\mathcal{H}}\left(\alpha - \frac{R_D^\beta \kappa}{K^N}\right) \right.$$

$$\left. + \left( \Gamma(2+N) - \Gamma\left(2+N, R_D (\lambda_w |\cos(\theta)| + \lambda_w |\sin(\theta)|)\right) \right) \mathcal{H}\left(\alpha - \frac{R_D^\beta \kappa}{K^N}\right) \right] d\theta. \tag{21}$$

In order to have an analytical expression for the integration over the variable $\theta$, we can express the the upper-incomplete Gamma functions of $\theta$ through their Taylor expansions, i.e.,

$$\Gamma\left(2+N, \left(\frac{\alpha K^N}{\kappa}\right)^{1/\beta} (\lambda_w |\cos(\theta)| + \lambda_w |\sin(\theta)|)\right) =$$
$$\Gamma(2+N) - \sum_{i=0}^\infty \frac{(-1)^i \left(\frac{\alpha K^N}{\kappa}\right)^{(N+2+i)/\beta}}{i!(i+N+2)} (\lambda_w |\cos(\theta)| + \lambda_w |\sin(\theta)|)^{(N+2+i)}, \tag{22}$$

and

$$\Gamma(2+N, R_D (\lambda_w |\cos(\theta)| + \lambda_w |\sin(\theta)|)) =$$
$$\Gamma(2+N) - \sum_{i=0}^\infty \frac{(-1)^i R_D^{(N+2+i)}}{i!(i+N+2)} (\lambda_w |\cos(\theta)| + \lambda_w |\sin(\theta)|)^{(N+2+i)}, \tag{23}$$

and, substituting (22) and (23) into (21), we get

$$\Lambda_N([0,\alpha)) = \frac{4\lambda_{PH}}{N!} \sum_{i=0}^\infty \frac{(-1)^i \lambda_w^{N+i}}{i!(N+i+2)} \left[ \frac{2^{\frac{N+i}{2}} \sqrt{\pi} \Gamma(\frac{N+i+1}{2})}{\Gamma(\frac{N+i+2}{2})} - \frac{\sqrt{2}\,_2F_1\left(\frac{1}{2}, \frac{N+i+1}{2}, \frac{N+i+3}{2}, \frac{1}{2}\right)}{N+i+1} \right]$$

$$\times \left[ R_D^{N+2+i} \; \mathcal{H}\left(\alpha - \frac{R_D^\beta \kappa}{K^N}\right) + \left(\frac{\alpha K^N}{\kappa}\right)^{\frac{N+i+2}{\beta}} \bar{\mathcal{H}}\left(\alpha - \frac{R_D^\beta \kappa}{K^N}\right) \right]. \tag{24}$$

Eventually, it worth recalling that the Heaviside function is defined as

$$\mathcal{H}(x) = \begin{cases} 1 & \text{if } x \geq 0 \\ 0 & \text{if } x < 0 \end{cases}$$

and that $\bar{\mathcal{H}}(x) = 1 - \mathcal{H}(x)$. Hence, by introducing the slack variable $\eta = N + i$ in (24) and defining the function $\chi_\eta(\lambda_w)$ as in (5), the proof is complete.

## APPENDIX B

### PROOF OF PROPOSITION 2

First, let note that the multi-user interference $\mathcal{I}_{MU}$ can be expressed as the sum of the interferences produced by the PHs belonging to $\Psi_N$, $N \in \{0, N_{max}\}$, where the processes $\Psi_N$ are assumed to be independent. Then, given the minimum propagation loss $L^{(0)}$, the characteristic function of $\mathcal{I}_{MU}$ is given by

$$\Phi\left(\omega; L^{(0)}\right) = \mathbb{E}\left(\exp\left\{j\omega \mathcal{I}_{MU}(L^{(0)})\right\}\right) = \prod_{N=0}^{N_{max}} \Phi_N\left(\omega; L^{(0)}\right). \tag{25}$$

where $\Phi_N\left(\omega; L^{(0)}\right)$ is the characteristic function of the interference generating from the PHs associated with $\Psi_N$. In order to compute $\Phi_N\left(\omega; L^{(0)}\right)$, we can invoke the Probability Generating Functional (PGFL) theorem for PPPs (see [9, Proposition 1.2.2]), thus getting

$$\Phi_N\left(\omega; L^{(0)}\right) = \exp\left(\mathbb{E}_{h^{(n)}}\left\{\int_{L^{(0)}}^{\infty}\left(\exp\left(j\omega h^{(n)}/\alpha\right) - 1\right)\widehat{\Lambda}_N([0,\alpha))d\alpha\right\}\right) \tag{26}$$

where $\widehat{\Lambda}_N([0,\alpha))$ is the first derivative of $\Lambda_N([0,\alpha))$ with respect to $\alpha$, given by

$$\widehat{\Lambda}_N([0,\alpha)) = \begin{cases} \frac{4\lambda_{PH}}{N!} \frac{K^N}{\kappa\beta} \sum_{\eta=N}^{\infty} \frac{(-1)^{\eta-N}\chi_\eta(\lambda_w)}{(\eta-N)!} \left(\frac{\alpha K^N}{\kappa}\right)^{\frac{\eta+2}{\beta}-1} & \text{if } \alpha < \frac{R_D^\beta \kappa}{K^N} \\ 0 & \text{if } \alpha \geq \frac{R_D^\beta \kappa}{K^N} \end{cases} \tag{27}$$

with $\chi_\eta(\lambda_w)$ defined as in (5).

Hence, similarly to what has been done in [10], we can use the notable result

$$\int_{\mathcal{N}}^{+\infty} (\exp\{j\omega\mathcal{A}/x\} - 1) x^{v-1} dx = (1/v)\mathcal{N}^v\left(1 - {}_1F_1(-v, 1-v, j\omega\mathcal{A}/\mathcal{N})\right)$$

and compute the expectation with respect to $h^{(n)}$ by taking advantage of the identity in [15, Eq. 7.521]. After some manipulations, the expression of the characteristic function is given by

$$\Phi_N\left(\omega; L^{(0)}\right) = \exp\left\{\frac{4\lambda_{PH}}{N!} \sum_{\eta=N}^{\infty} \frac{(-1)^{\eta-N}\chi_\eta\left(\lambda_w\right)}{(\eta-N)!(\eta+2)}\right.$$
$$\times \left[\left(\frac{L^{(0)}K^N}{\kappa}\right)^{\frac{\eta+2}{\beta}} \left(1 -_2F_1\left(1, -\frac{\eta+2}{\beta}, 1-\frac{(\eta+2)}{\beta}, \frac{j\omega}{L^{(0)}}\right)\right) \right.$$
$$\left.\left. - R_D^{\eta+2}\left(1 -_2F_1\left(1, -\frac{\eta+2}{\beta}, 1-\frac{(\eta+2)}{\beta}, \frac{j\omega K^N}{R_D^\beta \kappa}\right)\right)\right] \bar{\mathcal{H}}\left(L^{(0)} - \frac{R_D^\beta \kappa}{K^N}\right)\right\} \tag{28}$$

Then, invoking the Gil-Pelaez inversion theorem

$$F_{\mathcal{I}_{MU}}(z; L^{(0)}) = 1/2 - \int_0^\infty \frac{1}{\pi\omega} \operatorname{Im}\left\{e^{-j\omega z}\Phi\left(\omega; L^{(0)}\right)\right\} d\omega, \tag{29}$$

and recalling (9) and (25), Proposition 2 is finally proven.

## ACKNOWLEDGMENT

This work was supported by F.R.S.-FNRS under the EOS program (EOS project 30452698) and by INNOVIRIS under the COPINE-IOT project.